# Sinusoidal Displacement Describes Disorder in CsPbBr₃ Nanocrystal Superlattices

Umberto Filippi[1,2]*, Stefano Toso[3,4]*, Matheus G. Ferreira[3], Lorenzo Tallarini[3], Yurii P. Ivanov[1], Francesco Scattarella[5], Vahid Haghighat[6], Huaiyu Chen[7], Megan O. Hill Landberg[6], Giorgio Divitini[1], Jesper Wallentin[7], Cinzia Giannini[5]*, Liberato Manna[1]*, Dmitry Baranov[3]*

[1]Istituto Italiano di Tecnologia, Via Morego 30, 16163 Genova, Italy
[2]International Doctoral Program in Science, Università Cattolica del Sacro Cuore, Brescia 25121, Italy
[3]Division of Chemical Physics and NanoLund, Department of Chemistry, Lund University, P.O. Box 124, SE-221 00 Lund, Sweden
[4]Department of Chemical Engineering, Massachusetts Institute of Technology, Cambridge, Massachusetts 02139, United States
[5]Istituto di Cristallografia (CNR-IC), via Amendola 122/o, Bari 71025, Italy
[6]MAX IV Laboratory, Lund University, 22100 Lund, Sweden
[7]Synchrotron Radiation Research and NanoLund, Department of Physics, Lund University, 22100 Lund, Sweden

E-mail: umberto.filippi@iit.it, stefano.toso@chemphys.lu.se, cinzia.giannini@cnr.it, liberato.manna@iit.it, dmitry.baranov@chemphys.lu.se

## Abstract

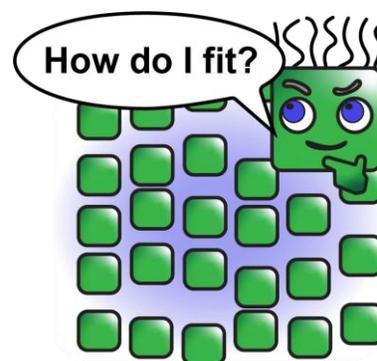

Disorder is an intrinsic feature of all solids, from crystals of atoms to superlattices of colloidal nanoparticles. Unlike atomic crystals, in nanocrystal superlattices a single misplaced particle can affect the positions of neighbors over long distances, leading to cumulative disorder. This elusive form of collective particle displacement leaves clear signatures in diffraction, but little is known about how it accumulates and propagates throughout the superlattice. Here we rationalize propagation and accumulation of disorder in a series of CsPbBr₃ nanocrystal superlattices by using synchrotron grazing incidence small- and wide-angle X-ray scattering. CsPbBr₃ nanocrystals of colloidal softness S in the range of 0.3-0.7 were obtained by preparing particles with different sizes and ligand mixtures, consisting of oleic acid and primary amines of variable lengths. Most diffraction patterns showed clear signatures of anisotropic disorder, with multilayer diffraction characteristics of high structural coherence visible only for the {100} axial directions and lost in all other directions. As the softness decreased, the superlattices transitioned to a more ordered regime where small-angle diffraction peaks became resolution-limited, and superlattice multilayer diffraction appeared for the (110) diagonal reflections. To rationalize these anisotropies in structural coherence and their dependence on superlattice softness, we propose a sinusoidal displacement model where longitudinal and transverse displacements modulate nanocrystal positions. The model explains experimental observations and advances the understanding of disorder in mesocrystalline systems as they approach the limits of structural perfection.

Colloidal nanocrystals are a compelling case of complex building blocks that can spontaneously self-organize into highly ordered three-dimensional solids (nanocrystal-based superlattices), and they attract intense interest because of emergent properties and applications in optoelectronics.[1-4] Nanocrystal-based superlattices exhibit disorder, owing to nanocrystal size and shape polydispersity, variations in ligand coverage, and weaker interparticle forces.[5] In diffraction experiments, this disorder manifests as peak broadening in higher order reflections. This is a signature of cumulative disorder, where displacement propagates through the structure as each particle position is influenced by that of the neighbors.[6, 7] This is the main reason why nanocrystal-based superlattices are studied by small-angle X-ray scattering, as superlattice interference is lost at wide angles.

However, colloidal perovskite nanocrystals with well-defined cubic shapes have been shown to assemble into superlattices with exceptional structural coherence, manifested by sharp and complex satellite peaks at wide angles.[8-14] These features suggest an intricate picture of disorder, in which precise superlattice periodicity and cumulative displacement can coexist. This raises the question of how positional correlations between neighboring nanocrystals propagate in different spatial directions within the superstructure, and, more generally, stimulates a discussion about the nature of disorder in nanoparticle assemblies. To address these questions, one needs to measure both small- and wide-angle X-ray scattering in multiple lattice directions on a series of high-quality superlattice samples with varying disorder. To this end, we performed experiments on perovskite nanocrystal superlattices at the ForMAX beamline of MAX IV, which allows simultaneous collection of grazing-incidence small- and wide-angle X-ray scattering (GISAXS/GIWAXS) patterns with high intensity and angular resolution. This setup provides complementary information, capturing both the average nanometric structure of superlattices (GISAXS) and the interference effects that reveal local structural coherence and disorder (GIWAXS).

Another piece of the puzzle comes from controlling the amount of disorder. Recently, some of us have reported that adopting mixed-ligand passivation enables control of disorder in $CsPbBr_3$ superlattices.[14] The mixed ligand passivation consists of oleic acid with aliphatic amines of variable length, where shorter amines lead to enhanced structural coherence in superlattices. Here, we build on this approach to tune the colloidal softness S of nanocrystals, defined as the ratio between the ligand shell thickness and the edge length of nanocrystals (S = $L/NC_{edge}$).[15-18] The studied softness range spans from S=0.3 (9.5 nm particles with oleic acid and octylamine) to S=0.7 (5.8 nm particles with oleic acid and oleylamine), and disorder increases with increasing S, as evidenced by broadening of the GISAXS spots (**Scheme 1**).

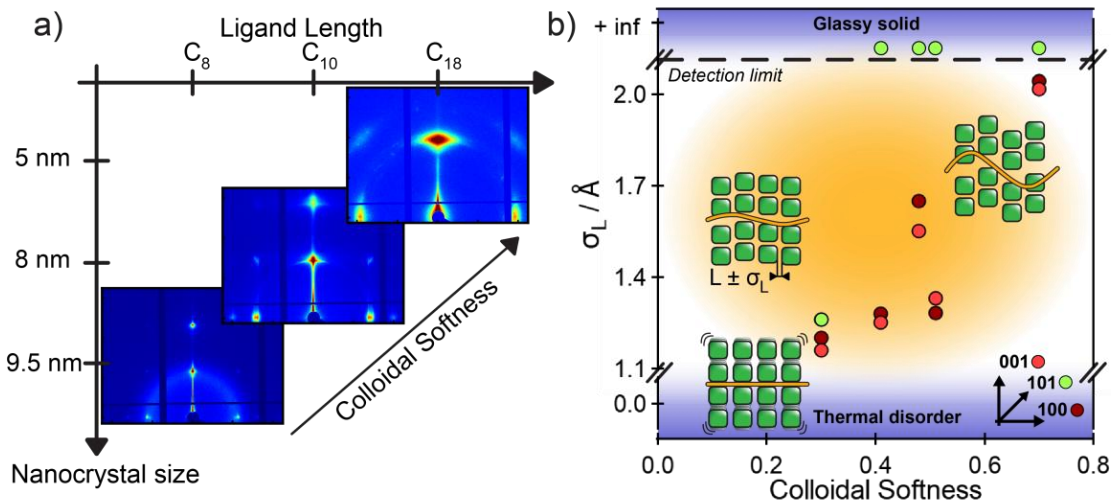

**Scheme 1.** **(a)** Synchrotron GISAXS patterns of three representative CsPbBr₃ nanocrystal superlattices as a function of ligand length and nanocrystal size, connected through colloidal softness (S). **(b)** Regimes of the structural disorder of nanocrystal superlattices as a function of the average nanocrystal longitudinal displacement ($\sigma_L$) and S. In the illustration, the $\sigma_L$=0 corresponds to the limit where nanocrystal positions are affected only by thermal-like (uncorrelated) fluctuations, and the $\sigma_L \rightarrow +\infty$ limit, a "superlattice catastrophe," corresponds to the case of a short-range order between nanocrystals, as in glassy solids.[19] The sinusoidal displacement model (wavy orange lines) captures the different disorder regimes through the amplitude, wavelength, and direction (transversal or longitudinal) of the oscillations of nanocrystal positions.

The experiments reported below provide evidence of high structural coherence for both in- and out-of-plane GIWAXS signals along the {100} directions of CsPbBr₃ superlattices. Except for the sample with the lowest S value (S=0.3), the collective X-ray interference is absent for other diffraction spots in the rest of the samples. In the S=0.3 sample, the signatures of satellite peaks were observed for the GIWAXS (101) reflection. These observations raise the paradox of superlattices that are highly ordered along the main lattice directions, and yet appear disordered along the directions given by their linear combinations. Such observations prompted us to develop a sinusoidal displacement model consisting of longitudinal and transversal components. The longitudinal component affects the face-to-face interparticle distance along the main lattice directions, and the transversal component affects the lateral displacement of nanocrystal rows and columns. The model explains the experimental evidence of anisotropic structural coherence while preserving the intrinsically cumulative nature of disorder in nanocrystal superlattices. This sinusoidal description of disorder in nanocrystal solids can likely be extended to other materials. The results deepen insight into the structure of nanocrystal assemblies and rationalize colloidal softness as a powerful tool for superlattice engineering.

## Results and Discussion

**Nanocrystals with Mixed Ligands.** The study was done on a series of superlattices grown from $CsPbBr_3$ nanocrystals capped with mixed ligands consisting of oleic acid and amines of different lengths: oleylamine ($C_{18}$), dodecylamine ($C_{12}$), decylamine ($C_{10}$) and octylamine ($C_8$).[14] The starting nanocrystals were synthesized using a common hot-injection method, reacting cesium oleate with $PbBr_2$ solubilized in a hot mixture of ligands and 1-octadecene, as summarized in the Methods. Representative high-resolution scanning transmission electron microscopy high-angle annular dark field (STEM-HAADF) images of the nanocrystals are shown in **Figures 1** and **S1** in the **Supporting Information** (SI). A shortening of the interparticle distance is noticeable, as expected from the incorporation of progressively shorter amines into the ligand shell of the nanocrystals. Upon drying concentrated dispersions on silicon or similar substrates (see Methods), the nanocrystals produce three-dimensional superlattices, with representative scanning electron microscopy (SEM) images of the tilted top view (8nm-$C_{18}$ nanocrystals) shown in **Figure 1e** and of a high-resolution region in **Figure 1f**. The edge lengths of the nanocrystals are approximately 8 nm for $C_{18}$, $C_{12}$, and $C_{10}$ ligands, and 9.5 nm for $C_8$ ligands, placing them in the intermediate quantum confinement regime. For comparison, a batch of quantum-confined nanocrystals with an edge length of 5.8 nm (sample 5nm-$C_{18}$) was synthesized by adapting the protocol reported by Dong *et al.*[20] with modifications.[21] **Figure 1g** shows the absorption spectra of dilute nanocrystal dispersions, illustrating the intermediate and strong confinement regimes. The colloidal softness was calculated for all samples using the nanocrystal edge length determined from TEM. The S values corresponding to the series of studied samples are 0.70 (5nm-$C_{18}$), 0.48 ($C_{18}$), 0.51 ($C_{12}$), 0.41 ($C_{10}$), and 0.3 ($C_8$). Similar values of softness were determined independently from the multilayer diffraction analyses (**Table 1**).

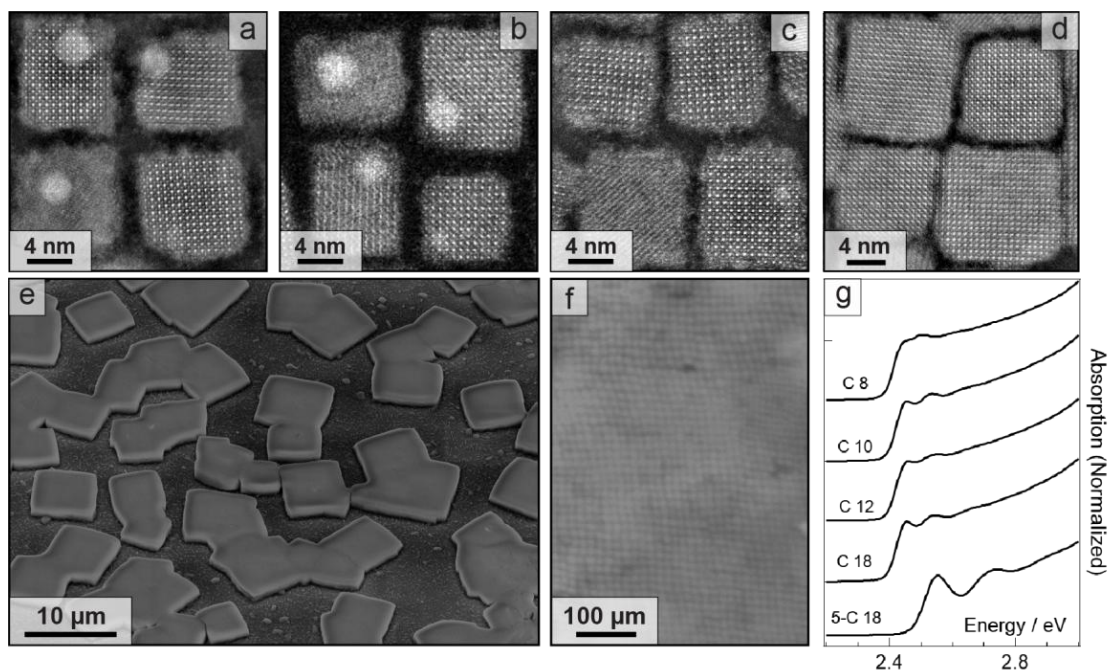

**Figure 1. Nanocrystals with mixed ligands.** (**a-d**) High-resolution STEM-HAADF images of CsPbBr$_3$ nanocrystals capped with oleic acid and amines of different lengths. From left to right: (**a**) oleic acid - oleylamine (8nm-C$_{18}$), (**b**) oleic acid - dodecylamine (C$_{12}$), (**c**) oleic acid - decylamine (C$_{10}$) and (**d**) oleic acid - octylamine (C$_8$). (**e, f**) Representative SEM images displaying a top view of 8nm-C$_{18}$ nanocrystal superlattices acquired with the substrate tilted 45° (e) and with no tilting (f). (**g**) Absorption spectra of toluene dispersions of nanocrystal samples.

**GISAXS Evidence of Cumulative Disorder.** The GISAXS and GIWAXS experiments described here were conducted at the ForMAX beamline[22] of the MAX IV synchrotron. Experimental details and data processing are described in the Methods section. **Figure 2a-f** displays the 2D GISAXS patterns and respective 1D slices of superlattice films grown from the dispersions of nanocrystals with mixed ligands shown in **Figure 1**. All GISAXS patterns exhibit well-defined spots, indicative of the nanoscale 3D periodicity of the superlattices. The GISAXS patterns were indexed as cubic space group $Pm\overline{3}m$ using SUNBIM4.0 software,[23] with a representative indexed pattern shown in **Figure 2d**. **Figure 2f** compares the in-plane (black curves) and out-of-plane (colored curves) 1D GISAXS profiles, from which the superlattice periodicities $\Lambda = 2\pi/\Delta q$ ($\Delta q = q_{n+1} - q_n$) were extracted (see **Table S1**). For 5nm-C$_{18}$, 8nm-C$_{18}$, C$_{12}$ and C$_{10}$ samples, identical periodicities in in-plane (x-y direction) and out-of-plane (z-direction) directions were obtained, consistent with nanocrystal sizes estimated from TEM and absorption spectra, once the ligands layer thickness is considered. For C$_8$, a small difference between the in-plane and out-of-plane periodicities (139 vs 137 Å) can be rationalized by residual strain after solvent evaporation.

The radial and azimuthal broadening of the first-order peaks (**Figure 2g**, see also **Figures S2, S3 and S4**) vary with the superlattice softness. Here, radial broadening carries information about nanoscale inhomogeneities in the superlattice periodicities and cumulative disorder, while azimuthal broadening indicates orientational disorder either at the level of whole superlattices or of multi-particle domains. Both radial and azimuthal broadening decrease from $C_{18}$ to $C_8$, as nanocrystals get bigger and ligands get shorter, indicating that a reduction in the nanocrystal softness increases the overall order of superlattices.

The GISAXS 2D patterns for all samples showed multiple spots, allowing us to compare how peak broadening changes with diffraction orders across different samples and use it as a qualitative distinction between cumulative and thermal-like nanocrystal displacements.[24-26] We observed that the peak width increases with the diffraction order for all samples except $C_8$ (as highlighted by arrows in **Figure 2f** and plotted in **Figure 2h**). In this case, the resolution-limited width in GISAXS resembles crystalline multilayer systems (*e.g.*, polyacene, amphiphilic crystals, and 2D Ruddlesden-Popper metal halides, see **Figure S5** for comparison).[27-29]

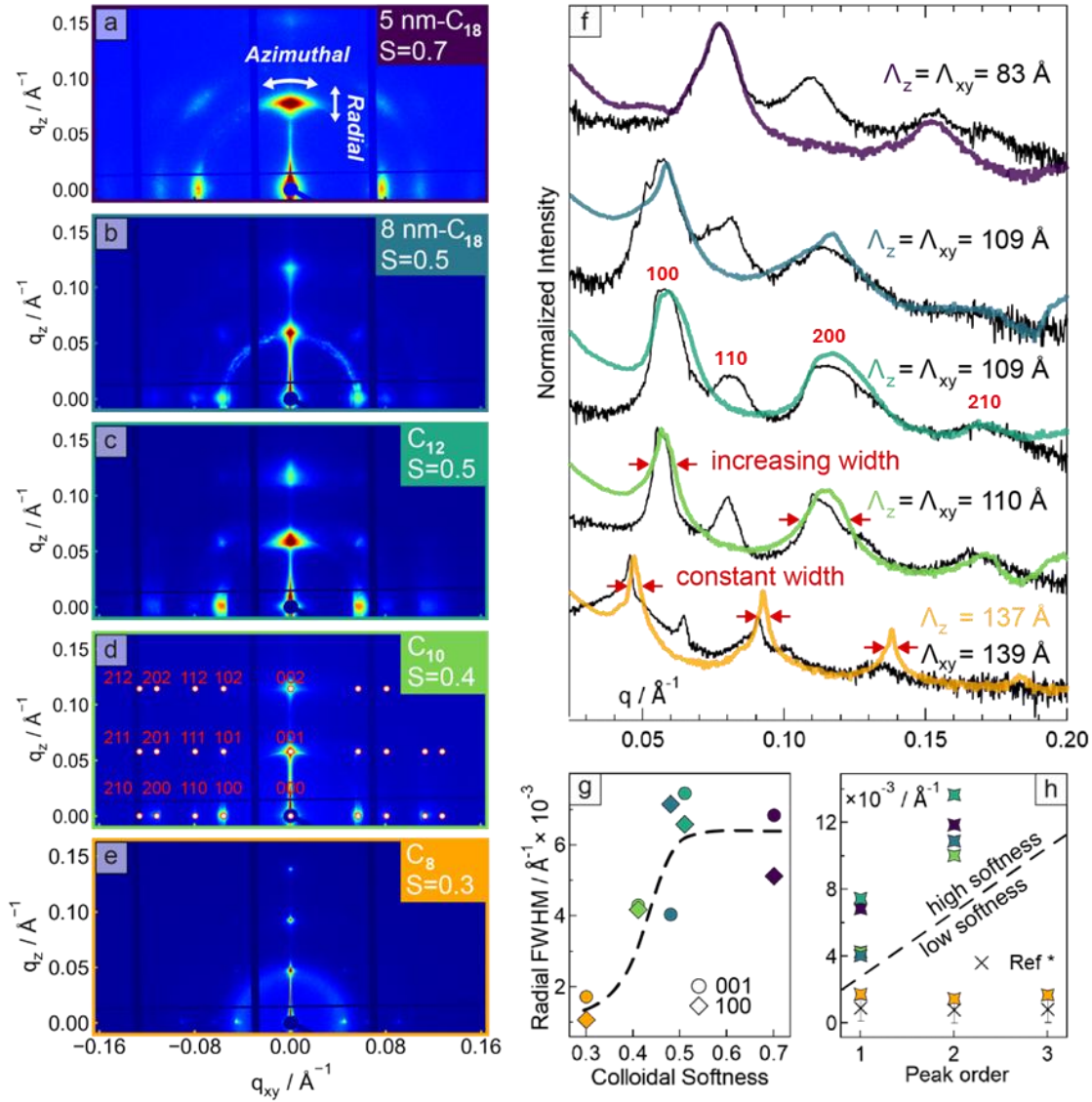

**Figure 2. GISAXS characterization of superlattices from nanocrystals capped with mixed ligands. (a-e)** Experimental GISAXS patterns from a series of nanocrystal superlattice samples, plotted on a logarithmic scale. From top to bottom: **(a)** 5nm-$C_{18}$ nanocrystals, **(b)** 8nm-$C_{18}$, **(c)** $C_{12}$, **(d)** $C_{10}$, and **(e)** $C_8$. Dark blue lines correspond to detector module gaps. **(f)** Intensity profiles extracted from GISAXS patterns. Colored profiles are taken along out-of-plane "z" direction. Underlying black profiles are taken along the in-plane "x-y" direction. Each profile is an average of five slices starting from the origin ($q_{xy} = q_z = 0$). For each superlattice profile, the extracted periodicities $\Lambda$ are reported ($\Lambda = 2\pi/\Delta q$, where $\Delta q = q_{n+1} - q_n$). The arrows highlight the increasing vs. constant peak width $\Delta q_z$ with diffraction order. **(g)** Radial broadening for in-plane (001) and out-of-plane (100) directions. **(h)** Out-of-plane GISAXS peak broadening as a function of the peak order. The literature example of an amphiphile thin film with a lamellar phase is reported with "×" and extracted from the reference.[28]

**GIWAXS Evidence of Disorder Anisotropy.** The as-synthesized $CsPbBr_3$ nanocrystals employed here as building blocks were established to have an orthorhombic structure (*Pbnm*, ICSD No. 98009-7851),[8] and GIWAXS patterns were indexed using a pseudocubic notation for simplicity, as clarified previously (**Figure 3a** shows a pattern with indexed spots relevant for the analysis).[8, 9] All GIWAXS patterns (**Figure 3a-e**) exhibit diffraction spots due to the fully-oriented nanocrystals inside the superlattice, with the (100) and (001) spots showing characteristic satellites due to the structural coherence.[8-13] To quantify cumulative disorder in multiple directions of the superlattice, we applied the multilayer diffraction routine developed for wide-angle $\theta$:2$\theta$ X-ray diffraction to GIWAXS data. The main results are summarized in **Table 1** and demonstrate shortening of the interparticle separation and increasing order with decreasing S. In addition to diffraction spots, the $C_8$ sample pattern shows arcs coming from residual non-assembled nanocrystals, which were subtracted for analysis (**Figures S7**, **S8** in the SI).

**Table 1.** Summary of GIWAXS multilayer diffraction fits. Parameters: d = nanocrystal lattice constant; L = interparticle distance (surface to surface); $\sigma_L$ = stacking disorder; N = nanocrystal thickness; $\sigma_N$ = nanocrystal thickness distribution. The average values are reported for brevity. For detailed fit parameters for each condition, see **Tables S2, S3**.

| Sample | d [Å] | L±$\sigma_L$ [Å] | N±$\sigma_N$ [planes] | S = L/(d·N) |
|---|---|---|---|---|
| **(001) peak** | | | | |
| 5nm-$C_{18}$ | | 36.5715±2.0435 | 8.8900±1.3030 | 0.70 |
| 8nm-$C_{18}$ | | 37.7075±1.584 | 13.3065±1.66 | 0.48 |
| $C_{12}$ | 5.884±0.020 | 37.9615±1.306 | 12.6265±1.5665 | 0.51 |
| $C_{10}$ | | 33.786±1.2635 | 14.0525±1.449 | 0.41 |
| $C_8$ | | 32.931±1.1815 | 18.7585±3.2575 | 0.30 |
| **(101) peak** | | | | |
| $C_{10}$ | 4.1059±0.0025 | 26.4771±1.8531 | ≈16 | - |
| $C_8$ | | 29.9941±1.2642 | ≈20 | - |

**Figure 3f** displays normalized diffraction intensity profiles for the axial (100)/(001) and the diagonal (101) peaks extracted from GIWAXS patterns. The multilayer diffraction fit parameters are summarized in **Table S2**, while the fits are shown in **Figure S6**. The interparticle spacing (L) and average nanocrystal displacement ($\sigma_L$) decrease with the shortening of the amine ligand (**Figure 3g**, **h**, respectively), consistent with GISAXS results discussed above and prior characterization using a laboratory diffractometer.[14] A significant finding is that, for each sample, $\sigma_L$ is comparable for (001) peak in-plane and out-of-plane dimensions within error, indicating that the nanocrystal displacement parameters follow the cubic symmetry of the superlattice CsPbBr₃ (**Figure 3h**). Slight differences in the atomic lattice constant and ligand interdigitation in the out-of-plane direction could be a further explanation of the difference in the two periodicities observed for $C_8$ superlattices in GISAXS.

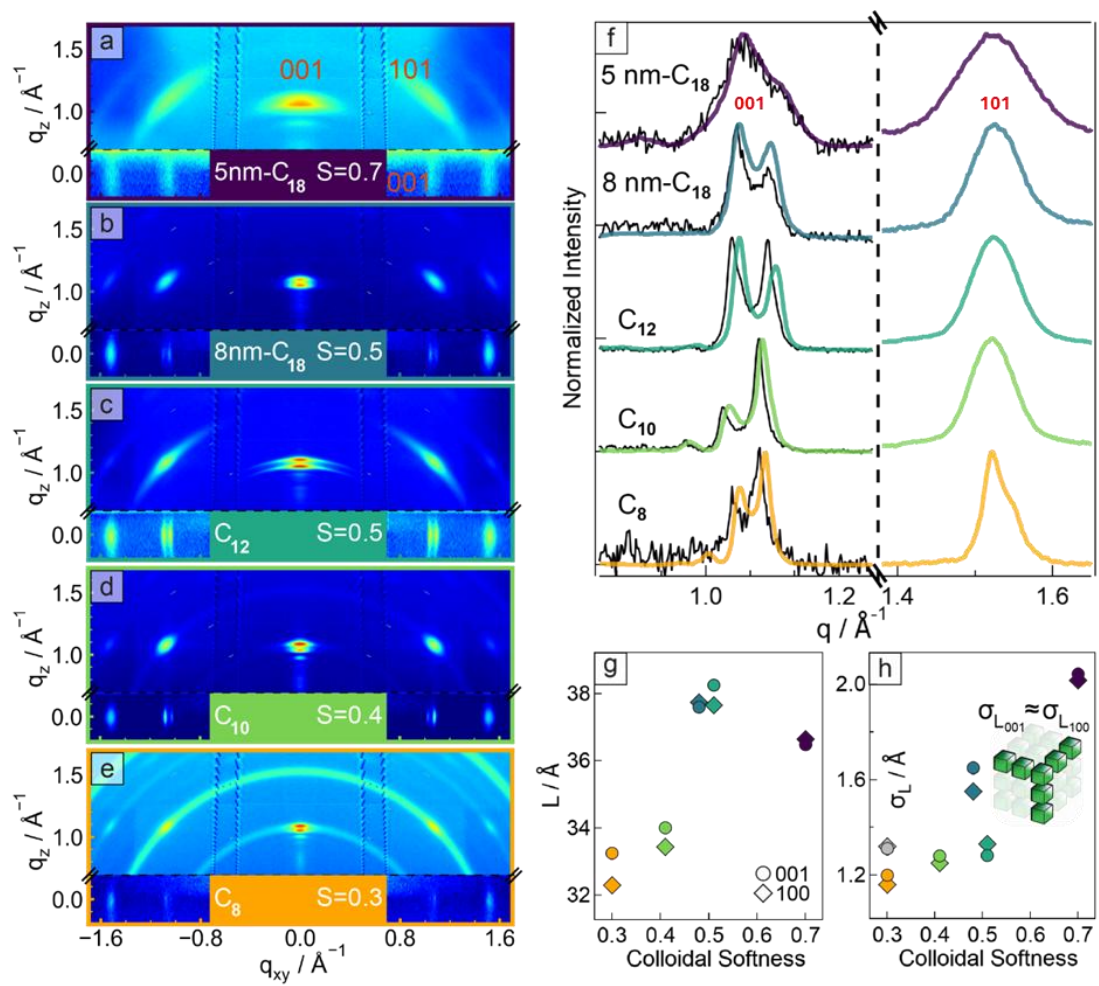

**Figure 3. GIWAXS characterization of superlattices from nanocrystals capped with mixed ligands. (a-e)** Experimental GIWAXS patterns from superlattice films of nanocrystals capped with mixed ligands. From top to bottom: **(a)** 5nm-$C_{18}$, **(b)** 8nm-$C_{18}$, **(c)** $C_{12}$, **(d)** $C_{10}$, and **(e)** $C_8$. **(f)** Intensity profiles extracted from GIWAXS patterns. Colored traces represent the out-of-plane (001) and (101) peak profiles, with the underlying black profile represents the in-plane (100) one. **(g, h)** Inter-nanocrystal distance (L) and average nanocrystal displacement ($\sigma_L$), respectively. For $C_8$ sample (S=0.3), grey points indicate fit values from the raw data before correction for the background of randomly oriented nanocrystals, and the orange points indicate fit values after correction. The cartoon inset in (h) depicts the isotropic order of nanocrystals along the three orthogonal directions.

**Diagonal Structural Coherence and the Need for a Different Model.** The initial observations of structural coherence in perovskite superlattices were satellites of the (100) and (200) Bragg reflections of the pseudo-cubic perovskite structure, which align to the main lattice directions of the assembly.[8, 9] Although (101) diffraction spots are visible in GIWAXS patterns (**Figure 3f**), they completely lacked such intensity modulation satellites for the $C_{18}$, $C_{12}$, and $C_{10}$ samples. This result contrasts with the 1D description of cumulative disorder implicit in the multilayer diffraction model, because collective interference should monotonically fade at higher $q$-values, q(002) > q(101). Indeed, multilayer diffraction simulations performed for the (101) peak using the same $\sigma_L$ value extracted for the (001) peak in **Figure 3h** indicate that collective interference modulations should be observed (see **Figure S9**). In contrast with this trend, the $C_8$ sample showed a clear X-ray scattering intensity modulation of the (101) peak. Notably, in this case, the $\sigma_L \approx 1.26$ Å value extracted from the (101) fit was compatible $\sigma_L$ value observed for the main axial directions. For comparison, the $C_{10}$ has lower $\sigma_L$ in the axial direction (*i.e.*, better structural coherence), but displays no interference modulation whatsoever for the (101) peak ($\sigma_L > 2$Å, estimated limit of detection).

We hypothesized that the absence of such features for the (101) planes arises from their sensitivity to combinations of multiple displacement directions as opposed to (100) planes. This difference comes from geometric considerations: the (100) planes lie parallel to the nanocube facets, and their alignment with neighboring nanocrystals is determined by translation along a single Cartesian axis. In contrast, (101) planes are inclined at 45° to the nanocube facets, so their alignment is coupled to two translational degrees of freedom. As a result, a displacement along one Cartesian axis disrupts only one of the three (100) families of planes but affects two of the three (101) families. Therefore, X-ray interference from (101) planes of nanocubes is more easily suppressed by a slight positional displacement of nanocrystals in the superlattice. Although $C_8$, $C_{10}$, and $C_{12}$ are all well-ordered superlattice samples from the (100) plane perspective, because of the clear presence of satellite peaks, high structural coherence in the {101} direction was detected only in $C_8$ . This can be rationalized by a loss in degrees of freedom (**Figure 4c**) with the increased rigidity of the nanocrystal packing as a consequence of low softness (S=0.3 for $C_8$, and 0.4 and 0.5 for $C_{10}$ and $C_{12}$, respectively). This interpretation ties with the higher peak sharpness observed in GISAXS for $C_8$ superlattices compared to all other investigated samples, which also suggests that such assemblies are less prone to rotation and shear-like movement than softer superlattices, and therefore $C_8$ is the most ordered superlattice either at the supra-crystal scale (GISAXS) and at the sub-crystal scale (GIWAXS).

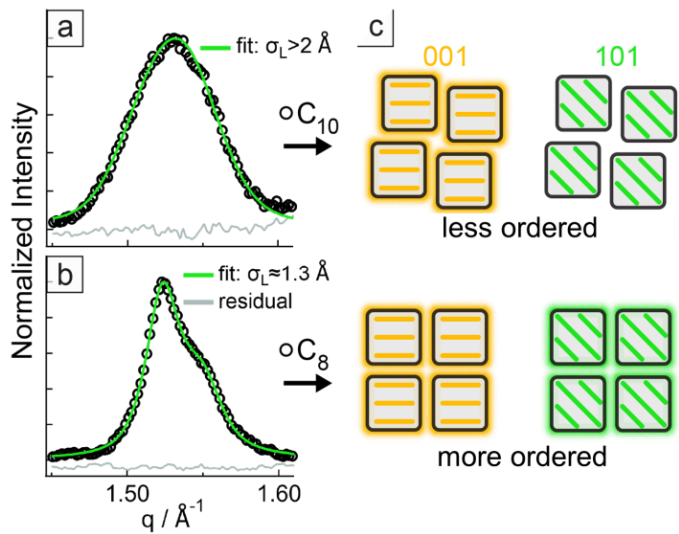

**Figure 4.** Experimental GIWAXS (101) peak profiles (circles, *ca.* $q_{center}$ = 1.53 Å$^{-1}$) and fits (continuous green lines) for **(a)** $C_{10}$ and **(b)** $C_8$ nanocrystal superlattice samples. The fit parameters are summarized in **Table 1**. **(c)** Illustration of the alignment (structural coherence) between (001) and (101) families of planes. In the case of more-ordered $C_8$ nanocrystals, the alignment is maintained for the (001) and (101) families of planes as highlighted by the glowing outlines. In the case of less-ordered $C_{10}$ nanocrystals, the alignment is preserved for (001) but lost for the (101) family of planes.

**A Sinusoidal Displacement Model.** The experimental findings suggest that nanocrystals with the lowest colloidal softness produce superlattices where structural coherence is better preserved in all directions of the lattice. That is consistent with prior observations of improved order in superlattices of large PbS nanocrystals[30] and colloidal $CsPbBr_3$ nanoplatelets with short octylamine ligands.[10] The observation of multilayer interference modulating the (101) reflections in GIWAXS data was inconsistent with a simplified model of cumulative disorder where random displacements were added to the interparticle spacing at each nanocrystal site and were sufficient to reproduce multilayer interference from the family of (100) planes.[13, 31] In other words, if the first and second order (h00) Bragg reflection were modulated, then the (101) reflection should be modulated too, contradicting experimental observations.

To rationalize this discrepancy, a displacement model is needed that satisfies several key requirements. It must decouple the average face-to-face interparticle distance variability (captured in the multilayer diffraction model by the parameter $\sigma_L$) from a shear-like lateral displacement. At the same time, the model should enforce correlation of various shifts across all the spatial directions, preserve cumulative behavior, and prevent the nanocrystal position from diverging away from the superlattice origin. To meet these criteria, we introduced a periodic displacement (u) model, in which the nanocrystal coordinates are shifted by longitudinal ($u_l$) and transversal ($u_t$) sinusoidal displacements from the ideal positions ($u_{ideal}$):

$$u = u_{ideal} + u_l + u_t \qquad \text{(eq. 1)}$$

Where:

$$u_{ideal} = \Lambda(i + j + k) \qquad \text{(eq. 2)}$$

$$u_l = A_l \left[ \sin\left( \frac{2\pi\Lambda i}{\lambda_l} + \phi \right) + \sin\left( \frac{2\pi\Lambda j}{\lambda_l} + \phi \right) + \sin\left( \frac{2\pi\Lambda k}{\lambda_l} + \phi \right) \right] \qquad \text{(eq. 3)}$$

$$u_t = A_t \left[ \sin\left( \frac{2\pi\Lambda k}{\lambda_t} + \phi \right) \cos\left( \frac{2\pi\Lambda j}{\lambda_t} + \phi \right) + \sin\left( \frac{2\pi\Lambda i}{\lambda_t} + \phi \right) \cos\left( \frac{2\pi\Lambda k}{\lambda_t} + \phi \right) + \sin\left( \frac{2\pi\Lambda j}{\lambda_t} + \phi \right) \cos\left( \frac{2\pi\Lambda k}{\lambda_t} + \phi \right) \right] \quad \text{(eq. 4)}$$

In the equations above, $\Lambda$ is the superlattice periodicity and $i, j, k$ are the coordinates of the *n*-th nanocrystal, $A_{l,t}$ and $\lambda_{l,t}$ are the amplitudes and wavelengths of the displacement, respectively, $\phi$ is a random phase shift. Longitudinal displacement alters the interparticle spacing along the edges and diagonals of the superlattice unit cell, leading to a similar broadening of collective interference features for both (001)/(002) axial and (101) diagonal peaks. Conversely, the transversal displacement affects the displacement of nanocrystals in the plane perpendicular to the wave propagation vector, that is, parallel to the facet of neighboring nanocrystals. **Figure 5c-e** shows the wide-angle X-ray diffraction patterns calculated using the previously-developed models[9, 10, 13] with modifications for periodic displacements of nanocrystal coordinates as described in eqs. 1-4, and either increasing $\lambda_t$ from 400 nm to 600 nm or decreasing $A_t$ from 3 nm to 5 nm. Both scenarios result in decreased disorder and lead to the multilayer diffraction of the (101) reflection, while the other peaks remain unchanged. The same result wouldn't have been achieved by tuning the parameters of the longitudinal displacement mode only (for example, by increasing $\lambda_l$ from 200 nm to 300 nm or decreasing $A_l$ from 0.3 nm to 0.2 nm, see **Figure S11**), which mostly affects the broadening of the (001) and (002) peaks. The main source of loss of structural coherence for the (101) reflection is the shear-like transversal oscillation, which shifts the nanocrystal perpendicular to the propagation of the displacement wave.

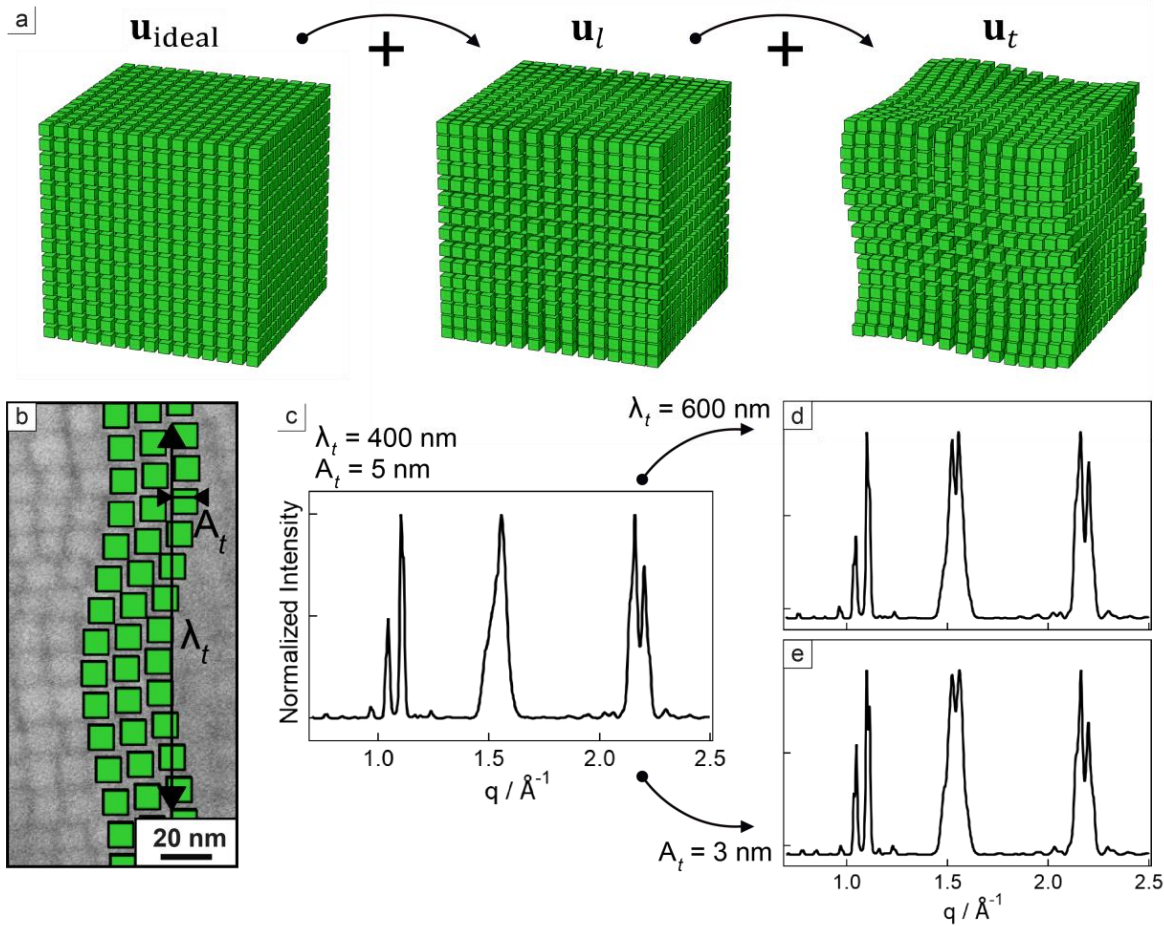

**Figure 5.** Periodic displacement in a nanocube superlattice. **(a)** Cartoon depicting the effect of the primary and secondary displacements on the packing of nanocrystals (eq. 1). **(b)** High-resolution SEM image of a superlattice fragment showing wavy nanocrystal packing with an overlay illustrating its amplitude ($A_t$) and wavelength ($\lambda_t$). **(c-e)** Calculated wide-angle X-ray diffraction patterns of CsPbBr$_3$ nanocrystal superlattices showing effect of changing $\lambda_t$ and $A_t$ onto multilayer interference of (101) peak.

We hypothesized that $A_t$ is minimized in the C$_8$ sample (S=0.3) because large nanocrystals and short interparticle spacing restrict nanocrystal movement, dampening the amplitude of the periodic displacement. In contrast, softer nanocrystals (e.g., S=0.4−0.7 for C$_{10}$, C$_{12}$, C$_{18}$, and 5-C$_{18}$ samples) exhibit more flexibility and thus erased {101} multilayer interference. Another interpretation is to view $\mathbf{u}_l$ and $\mathbf{u}_t$ as frozen waves that statically displace nanocrystals from their equilibrium positions. In this analogy, $\mathbf{u}_l$ and $\mathbf{u}_t$ resemble static acoustic phonon modes, with displacements parallel and perpendicular to the propagation direction, respectively. Thus, reduced nanocrystal softness corresponds to a longer spatial period of oscillation, analogous to the decrease in frequency of a harmonic oscillator with increasing reduced mass.[32]

Wave-like displacements caused by various mechanisms can also be observed in other crystalline systems, such as martensitic phases,[33] liquid crystals,[34, 35] and metal alloys.[36, 37] In liquid crystals, for example, twisted nematic or smectic structures, wave-like or helical, emerge due to molecular anisotropy, elasticity, or chirality, resulting in a periodic alignment of molecules.[38] The wave-like nanocrystal packing in superlattices has been observed in SEM and TEM images in this work (see **Figure 5b**) as well as across the literature.[30, 32, 39-44] It is therefore plausible that, similarly to other ordered systems, ligand-ligand interactions and structural inhomogeneities in nanocrystal assemblies could give rise to periodic displacements.

## Conclusions

This study demonstrates that perovskite $CsPbBr_3$ nanocrystals capped with mixed ligands can self-assemble into superlattices exhibiting high structural coherence in both in-plane and out-of-plane directions. Through combining GISAXS and GIWAXS synchrotron measurements, we showed that the structural order in these superlattices systematically depends on colloidal softness, a tunable parameter defined by the ratio between ligand shell thickness and nanocrystal edge length. A counterintuitive observation of a broadening at wide angles (consistent with cumulative disorder) but resolution-limited peak broadening at small angles (suggesting an apparent thermal-like, uncorrelated disorder) in the nanocrystals with S=0.3 prompted a deeper explanation. To rationalize these observations, we deconstructed nanocrystal displacements into longitudinal and transversal components, introducing sinusoidal modulation of nanocrystal coordinates that reconciles observed diffraction patterns and bridges regimes of thermal-like (where $A \to 0$ and $\lambda \to \infty$) and powder-like uncorrelated disorder (where $A \to \infty$ and $\lambda \to 0$). The model also offers a physical analogy to static acoustic phonon modes, linking decreased nanocrystal softness to reduced displacement amplitude and longer spatial modulation periods. Such a wave-like deformation closely resembles the periodic structural modulations found in liquid crystals,[33-35, 45] reinforcing the analogy between nanocrystal superlattices and soft condensed matter systems. The periodic nanocrystal displacements invite compelling analogies with phonon modes and atomic vibrations. As their exploration as quantum materials continues across nanoscience, this highlights the potential of nanocrystal superlattices as model systems for investigating excitonic and vibrational interactions[46-48] and excitation transport.[4, 49-51]

## Methods

*Nanocrystal synthesis and characterization.* Samples of 8-10 nm $CsPbBr_3$ nanocrystals passivated with mixed ligands (8nm-$C_{18}$, $C_{12}$, $C_{10}$, and $C_8$ series) were synthesized by hot-injection following a previously reported procedure with minor variations.[14] Briefly, the syntheses consisted of injecting 0.5 ml of 0.073 M cesium oleate solution in 1-octadecene into the hot lead bromide (*ca.* 0.036 M $PbBr_2$) solubilized in a mixture of alkylamine, oleic acid, and 1-octadecene. The temperature of the mixture was different for different amines (160 °C for 8nm-$C_{18}$, and 170 °C for $C_8$, $C_{10}$, and $C_{12}$ amines). The as-synthesized nanocrystals were isolated either by centrifugation alone ($C_{18}$ and $C_{12}$) or with the aid of anti-solvent ethyl acetate ($C_{10}$ and $C_8$). After isolation, the precipitate was re-dispersed in 300 µL toluene, forming a concentrated dispersion, centrifuged to remove undissolved material, and the obtained supernatant was used for subsequent characterization and superlattice growth. The quantum-confined nanocrystals of $CsPbBr_3$ (5nm-$C_{18}$) were synthesized using the hot-injection synthesis of Dong *et al.*,[20] with the detailed procedure reported by Gomes Ferreira *et al.*[21] UV-Vis optical absorption spectra were recorded on dilute dispersions of nanocrystals in toluene using Cary 500 ($C_8$, $C_{10}$, $C_{12}$, and 8nm-$C_{18}$ samples) and Perkin Elmer Lambda 1050 (5nm-$C_{18}$ sample) spectrometers. High-resolution scanning transmission electron microscopy (HRSTEM) images were acquired on a probe-corrected Thermo Fisher Spectra 30-300 STEM operated at 300 kV. Images were acquired on a High-Angle annular Dark Field (HAADF) detector with a current of 50 pA. SEM imaging was performed on a superlattice film grown on top of a silicon substrate, using a field-emission SEM JEOL JSM-7500 FA operating at an acceleration voltage of 25 kV.

*Synchrotron experiments and data analysis.* A thin film of nanocrystal superlattices for synchrotron experiments was prepared by drop-casting nanocrystal dispersions in toluene on silicon nitride substrates (1000 nm-thick SiN, Silson, product code M1000143, SiRN-5.0-200-2.5-1000) and leaving them to dry slowly, enclosed in a Petri dish. The nanocrystal concentration in the drop-cast dispersion was approximately 0.5 µM, to allow the growth of isolated superlattices. The concentrated nanocrystal dispersions obtained as described above, are stable over the course of 7-10 days needed for shipping and handling at the synchrotron, and do not show visible signs of degradation during handling under ambient conditions. GISAXS and GIWAXS experiments were performed using the ForMAX beamline at the MAX IV Laboratory.[22] For GISAXS, ForMAX employs an Eiger2 4M detector in an 8-meter vacuum vessel, while GIWAXS uses a custom windmill-shaped detector with a central aperture for GISAXS transmission. In the GISAXS/GIWAXS experiments, incident angles of 1.400° and 4.237° were used with a 14.31 keV beam, providing approximately $5.5 \cdot 10^{14}$ photons/s total flux. The 4.237° incidence angle was employed to collect the 2D map from which the 101 peak profiles were extracted and reported in **Figure 4** for $C_{10}$ and $C_8$ superlattices and in **Figure S10** for all other superlattices. The beam size was $50 \times 50$ µm² at normal incidence, which corresponds to the elongated footprint of the beam on the sample with a length of *ca.* 2 mm for 1.400° incident angle and *ca.* 680 µm for 4.237° incident angle. The sample-to-detector distances were 2990 mm for GISAXS and 142 mm for GIWAXS. Data calibration for GISAXS and GIWAXS was performed by employing silver behenate and $LaB_6$ powders as standards, respectively. The patterns were indexed using SUNBIM4.0 software.[23] The 1D profiles were extracted from the 2D maps by averaging 5 pixel lines in correspondence with the desired spots.

## Acknowledgements

We thank Simone Lauciello (Electron Microscopy Facility at IIT) for help with HRSEM analysis of superlattices and Kim Nygård (ForMAX beamline, MAX IV) for assistance with experiments and discussions. We acknowledge the MAX IV Laboratory for beamtime on the ForMAX beamline under proposal 20230363. Research conducted at MAX IV, a Swedish national user facility, is supported by Vetenskapsrådet (Swedish Research Council, VR) under contract 2018-07152, Vinnova (Swedish Governmental Agency for Innovation Systems) under contract 2018-04969 and Formas under contract 2019-02496. S. T. acknowledges the European Union's Horizon Europe research and innovation programme under the Marie Skłodowska-Curie Funding Program (Project SUPER-QD, Grant Agreement No. 101148934). The work of H.C. and J.W. received support from the Swedish Research Council (contract no. 2021-04273), the Olle Engkvist Foundation and the Essence project. This project received funding from the European Research Council (ERC) under the European Union's Horizon 2020 research and innovation program (Grant 801847). The work of L.T., M.G.F., and D.B. was funded by the European Union (ERC Starting Grant PROMETHEUS, project no. 101039683). Views and opinions expressed are, however, those of the authors only and do not necessarily reflect those of the European Union or the European Research Council Executive Agency. Neither the European Union nor the granting authority can be held responsible for them. D.B. acknowledges partial support from the grant NSFPHY-2309135 to the Kavli Institute for Theoretical Physics for participation in the KITP Program on Nanoparticle Assemblies: A New Form of Matter with Classical Structure and Quantum Function, that stimulated work on the MAX IV beamtime proposal.

## Supporting Information Available

Electron microscopy characterization, summaries of multilayer diffraction fits, GISAXS and GIWAXS analyses, calculated diffraction patterns.

## CRediT Statement

Umberto Filippi: Conceptualization (equal), Data curation (lead), Formal analysis (lead), Investigation (equal), Methodology (equal), Software (equal), Validation (lead), Visualization (supporting), Writing – Original Draft Preparation (equal), Writing – review and editing (equal).

Stefano Toso: Conceptualization (equal), Data curation (equal), Formal analysis (equal), Investigation (equal), Methodology (equal), Software (lead), Validation (equal), Visualization (lead), Writing – Original Draft Preparation (equal), Writing – review and editing (equal).

Matheus G. Ferreira: Data Curation (supporting), Investigation (supporting), Resources (supporting);

Lorenzo Tallarini: Data Curation (supporting), Investigation (supporting);

Yurii P. Ivanov: Data curation (lead), Formal analysis (equal), Investigation (equal), Visualization (equal), Writing – review and editing (supporting);

Francesco Scattarella: Data curation (supporting), Formal analysis (supporting), Investigation (supporting), Visualization (supporting);

Vahid Haghighat: Investigation (supporting), Writing – review and editing (supporting);

Huaiyu Chen: Investigation (supporting), Software (supporting), Writing – review and editing (supporting);

Megan O. Hill Landberg: Investigation (supporting), Software (supporting),

Giorgio Divitini: Funding acquisition (supporting), Project administration (supporting), Resources (supporting), Supervision (supporting), Writing – review and editing (supporting);

Cinzia Giannini: Investigation (supporting), Resources (supporting), Supervision (supporting), Writing – review and editing (supporting).

Jesper Wallentin: Data Curation (supporting), Formal analysis (supporting), Resources (supporting), Software (Supporting), Writing – review and editing (supporting).

Liberato Manna: Investigation (supporting), Resources (supporting), Supervision (supporting), Writing – review and editing (supporting).

Dmitry Baranov: Conceptualization (equal), Data curation (supporting), Formal analysis (supporting), Funding acquisition (lead), Investigation (equal), Methodology (equal), Project administration (lead), Resources (lead), Software (supporting), Supervision (lead), Validation (equal), Visualization (supporting), Writing – Original Draft Preparation (equal), Writing – review and editing (equal).

*Supporting Information for*

**Sinusoidal Displacement Describes Disorder in CsPbBr$_3$ Nanocrystal**

**Superlattices**

Umberto Filippi[1,2]*, Stefano Toso[3,4]*, Matheus G. Ferreira[3], Lorenzo Tallarini[3], Yurii P. Ivanov[1], Francesco Scattarella[5], Vahid Haghighat[6], Huaiyu Chen[7], Megan O. Hill Landberg[6], Giorgio Divitini[1], Jesper Wallentin[7], Cinzia Giannini[5]*, Liberato Manna[1]*, Dmitry Baranov[3]*

[1]Istituto Italiano di Tecnologia, Via Morego 30, 16163 Genova, Italy
[2]International Doctoral Program in Science, Università Cattolica del Sacro Cuore, Brescia 25121, Italy
[3]Division of Chemical Physics and NanoLund, Department of Chemistry, Lund University, P.O. Box 124, SE-221 00 Lund, Sweden
[4]Department of Chemical Engineering, Massachusetts Institute of Technology, Cambridge, Massachusetts 02139, United States
[5]Istituto di Cristallografia (CNR-IC), via Amendola 122/o, Bari 71025, Italy
[6]MAX IV Laboratory, Lund University, 22100 Lund, Sweden
[7]Synchrotron Radiation Research and NanoLund, Department of Physics, Lund University, 22100 Lund, Sweden

E-mail: umberto.filippi@iit.it, stefano.toso@chemphys.lu.se, cinzia.giannini@cnr.it, liberato.manna@iit.it, dmitry.baranov@chemphys.lu.se

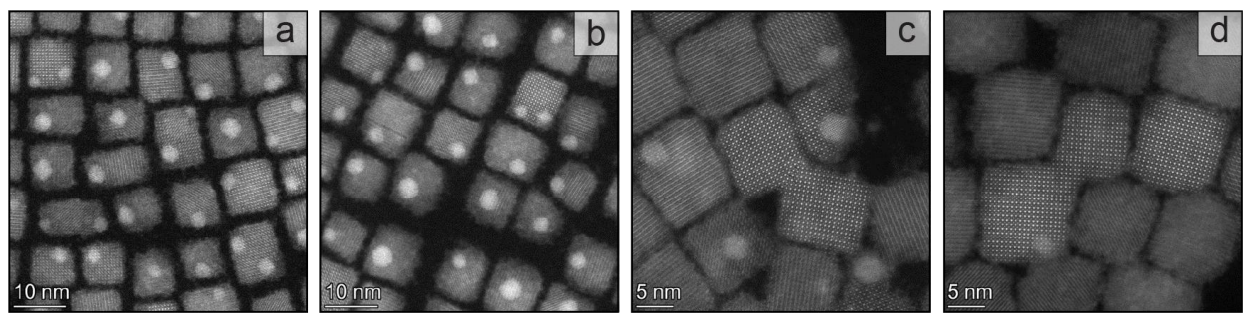

**Figure S1.** High-resolution STEM-HAADF (High-Angle Annular Dark Field) images of monolayers of (a),(b) 8nm-C$_{18}$ nanocrystals (well-separated from each other) and (c),(d) C$_8$ nanocrystals (some necking occurs under the imaging conditions).

**Table S1.** Superlattice periodicities from the GISAXS profiles (Figure 2f in the main text).

| Sample | Periodicity [Å] | | Crystal System | Space group |
|---|---|---|---|---|
| | Out-of-plane | In-plane | | |
| C$_{18}$ 5 nm | 82.8 ± 0.8 | | Cubic | Pm-3m |
| C$_{18}$ 8 nm | 109.1±1.3 | | Cubic | Pm-3m |
| C$_{12}$ | 108.6±1.2 | | Cubic | Pm-3m |
| C$_{10}$ | 109.9±1.2 | | Cubic | Pm-3m |
| C$_8$ | 136.8 ± 0.3 | 139.1 ± 0.3 | Cubic | Pm-3m |

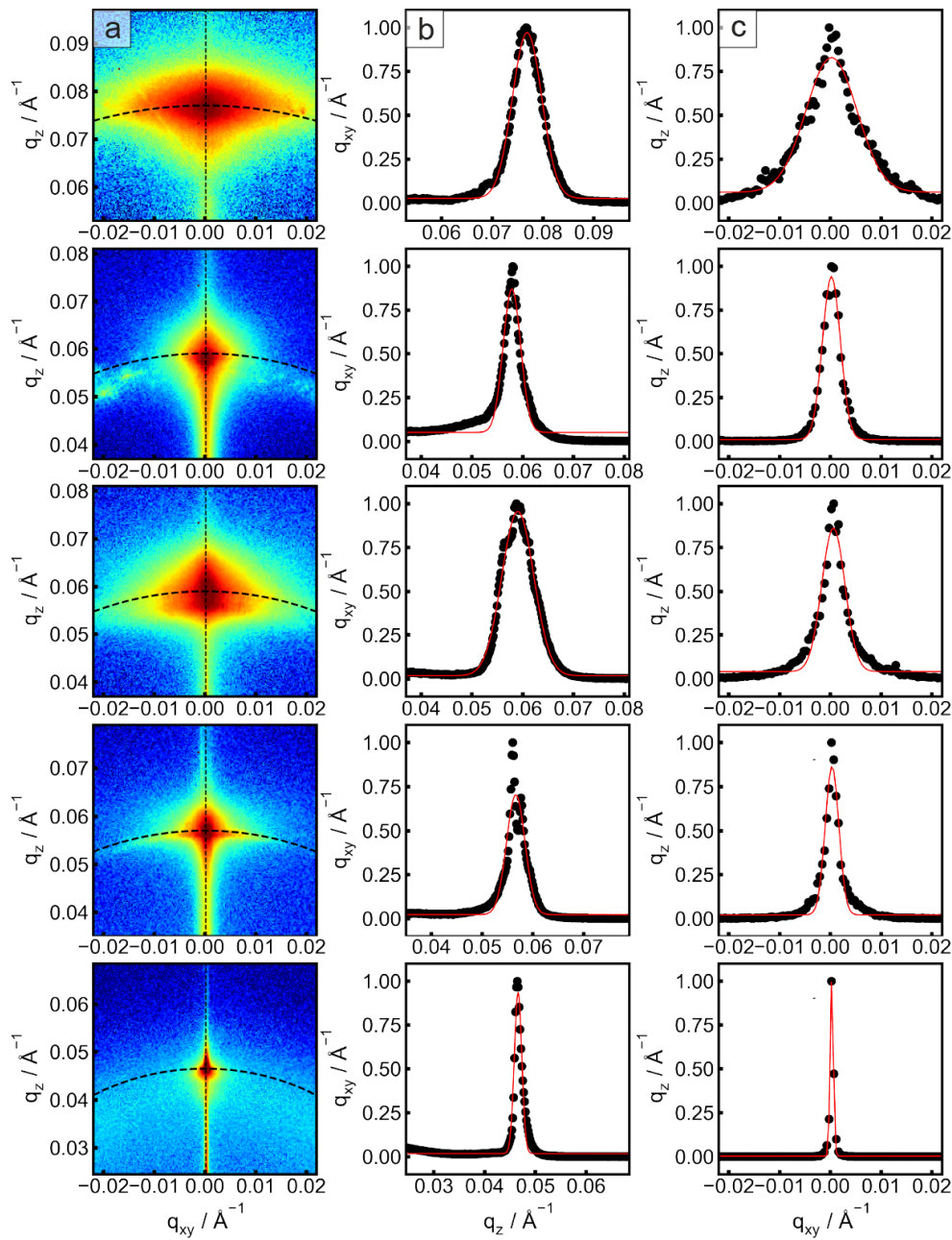

**Figure S2.** (a) 001 GISAXS peak, (b) azimuthal broadening profile and Gaussian fitting, (c) radial broadening profile and Gaussian fitting. From top to bottom: 5nm-$C_{18}$, 8nm-$C_{18}$, $C_{12}$, $C_{10}$, and $C_8$ nanocrystal superlattice samples.

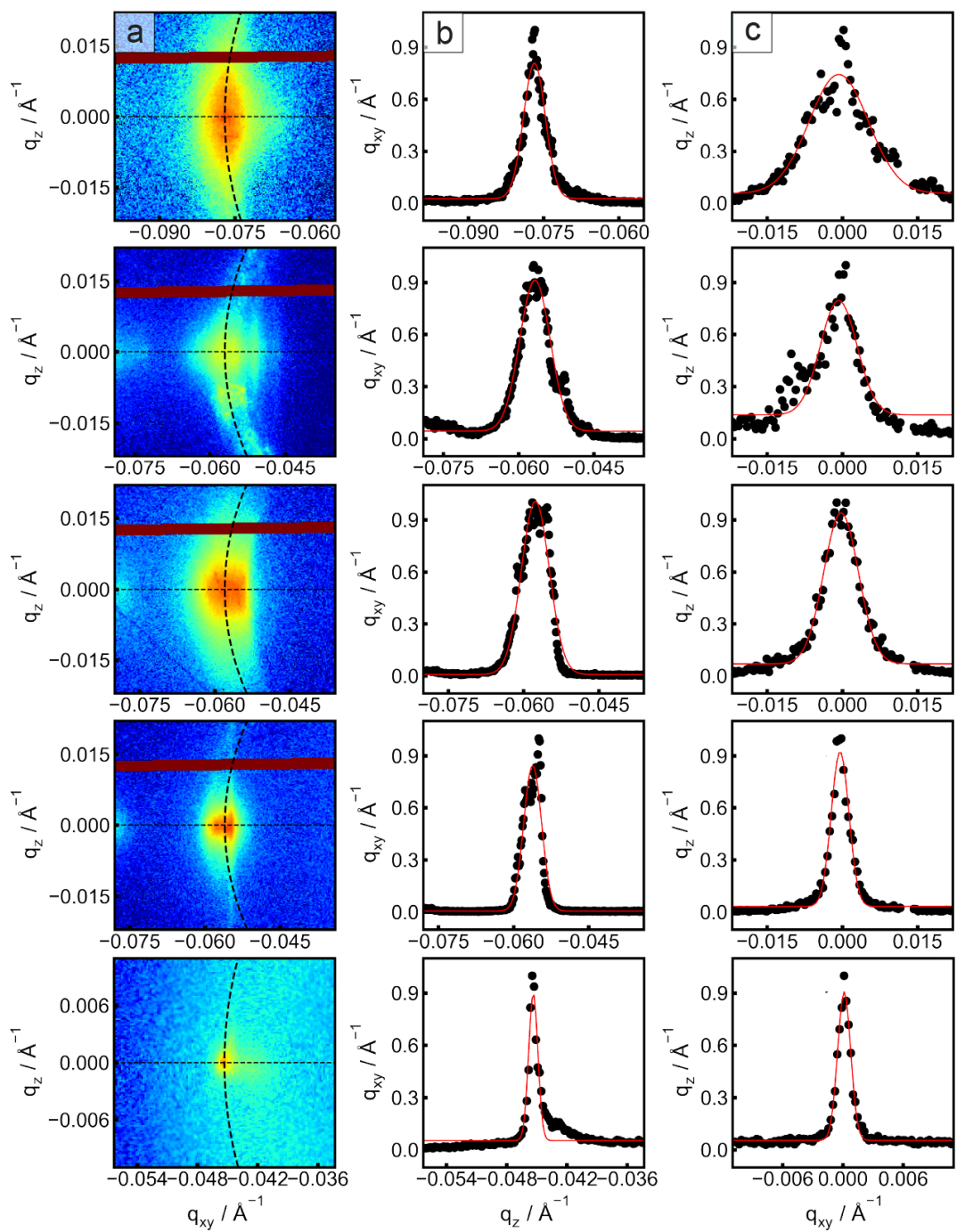

**Figure S3.** (a) 100 GISAXS peak, (b) azimuthal broadening profile and Gaussian fitting, (c) radial broadening profile and Gaussian fitting. From top to bottom: 5nm-$C_{18}$, 8nm-$C_{18}$, $C_{12}$, $C_{10}$, and $C_8$ nanocrystal superlattice samples.

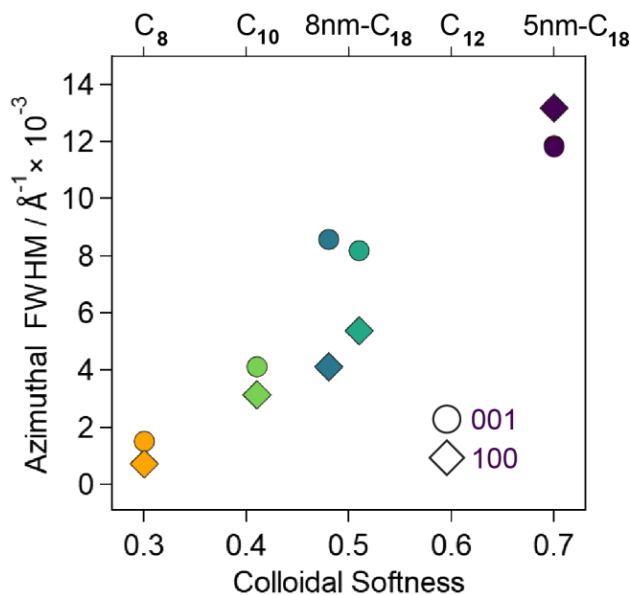

**Figure S4.** GISAXS azimuthal broadening for in-plane (001) and out-of-plane (100) directions.

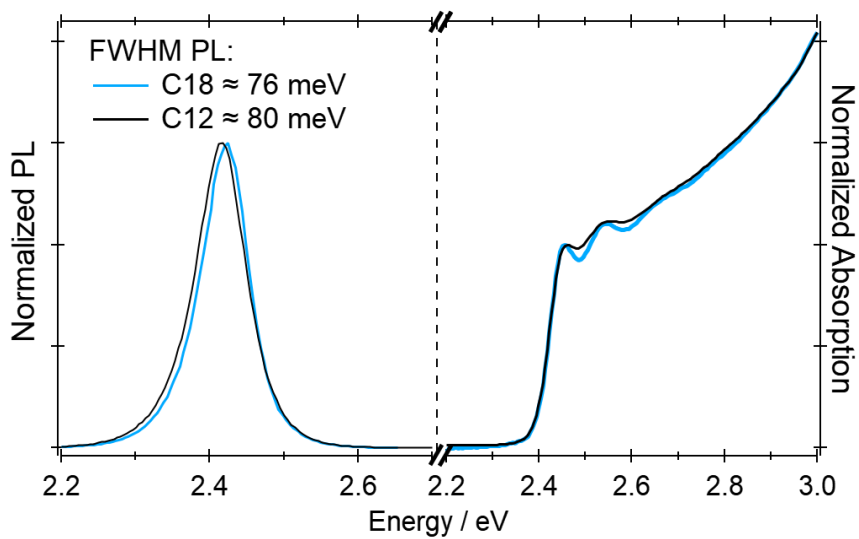

**Figure S5.** Comparison of the photoluminescence (PL) and absorption spectrum of $C_{12}$ and $C_{18}$ nanocrystals dispersed in toluene. The wider PL spectrum and less pronounced excitonic absorption peaks indicate a larger size dispersion for $C_{12}$ nanocrystals than for 8nm-$C_{18}$ nanocrystals.

**Table S2. Multilayer diffraction fit results of GIWAXS patterns.** Structure parameters extracted from the multilayer diffraction fits. The fits are shown in Figure S6. To account for the disordered contribution present in the $C_8$ GIWAXS pattern (Figure 3e), we subtracted a Gaussian background from the peak profile; the fitting results before and after subtraction are reported in *C8 and C8 rows, respectively. Parameters: d = nanocrystal lattice constant; L = interparticle distance (surface to surface); $\sigma_L$ = stacking disorder; N = nanocrystal thickness; $\sigma_N$ = nanocrystal thickness distribution, S – colloidal softness [calculated as L/(d·N)].

| Sample | d [Å] | L [Å] | $\sigma_L$ [Å] | N [planes] | $\sigma_N$ [planes] | S = L/(d·N) |
|---|---|---|---|---|---|---|
| **(100) Out-of-plane** | | | | | | |
| **5nm-C$_{18}$** | 5.886 | 36.493 | 2.041 | 8.879 | 1.239 | 0.70 |
| **8nm-C$_{18}$** | 5.866 | 37.814 | 1.567 | 13.531 | 1.561 | 0.48 |
| **C$_{12}$** | 5.852 | 37.661 | 1.282 | 12.249 | 2.129 | 0.53 |
| **C$_{10}$** | 5.866 | 33.567 | 1.281 | 13.938 | 2.012 | 0.41 |
| **C$_8$** | 5.860 | 32.677 | 1.203 | 19.116 | 3.501 | 0.29 |
| ***C$_8$** | 5.858 | 32.677 | 1.325 | 19.212 | 2.872 | 0.29 |
| **(100) In-plane** | | | | | | |
| **5nm-C$_{18}$** | 5.906 | 36.650 | 2.046 | 8.901 | 1.367 | 0.70 |
| **8nm-C$_{18}$** | 5.910 | 37.601 | 1.601 | 13.082 | 1.759 | 0.49 |
| **C$_{12}$** | 5.898 | 38.262 | 1.330 | 13.004 | 1.004 | 0.50 |
| **C$_{10}$** | 5.891 | 34.005 | 1.246 | 14.167 | 0.886 | 0.41 |
| **C$_8$** | 5.905 | 33.185 | 1.160 | 18.401 | 3.014 | 0.31 |
| ***C$_8$** | 5.909 | 33.126 | 1.31 | 18.343 | 3.504 | 0.31 |

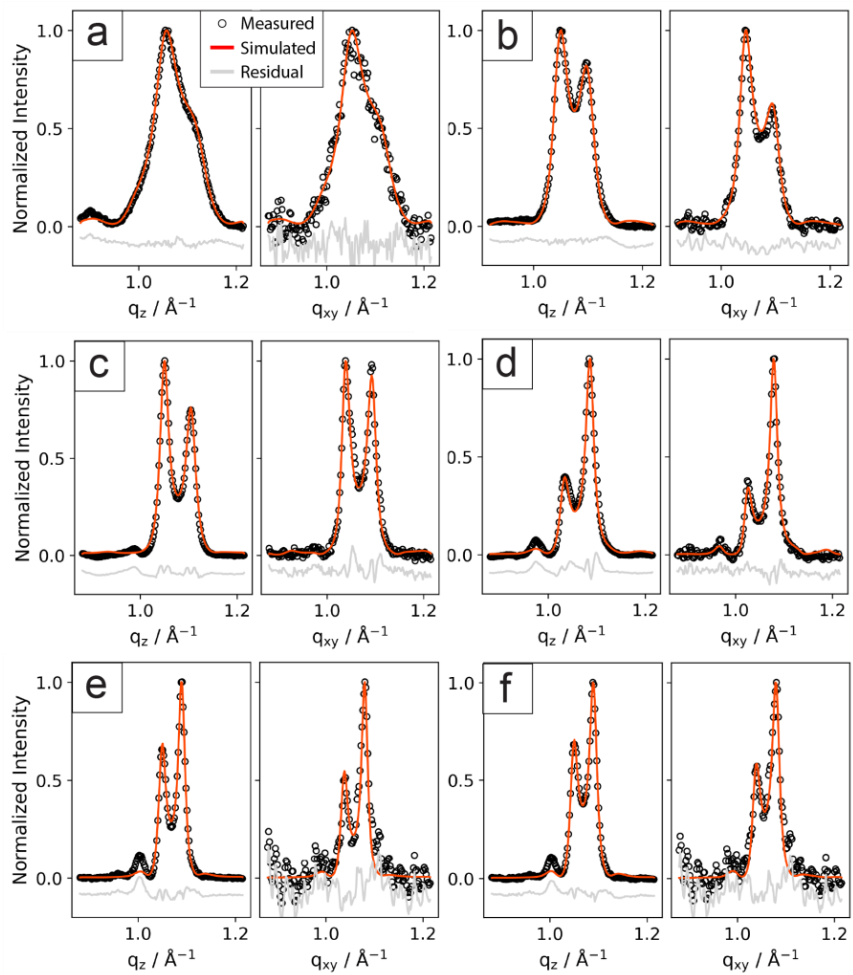

**Figure S6. Multilayer diffraction fits of GIWAXS patterns.** a-f) Out-of-plane (100) (left) and in-plane (100) (right) diffraction profiles extracted from the 2D GIWAXS patterns reported in Figure 3. They correspond to (a) 5nm-$C_{18}$, (b) 8nm-$C_{18}$, (c) $C_{12}$, (d) $C_{10}$ and (e) $C_8$, and (f) *$C_8$.

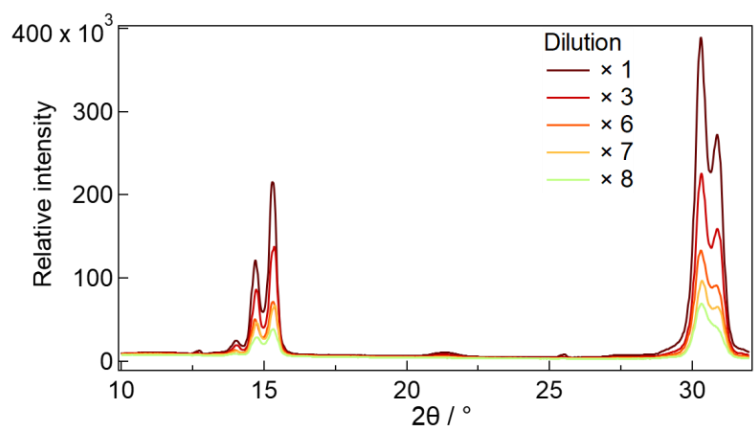

**Figure S7.** Concentration-dependent $C_8$ superlattice order.

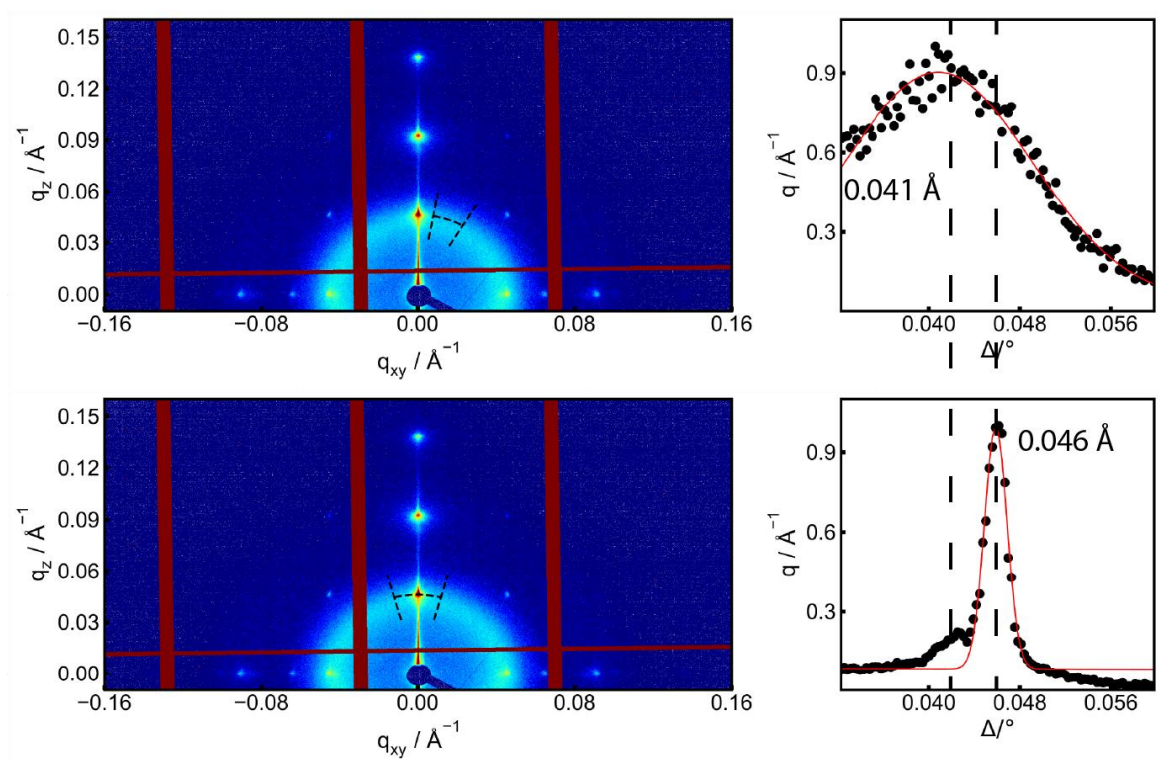

**Figure S8.** GISAXS C$_8$ fitting of the disordered component.

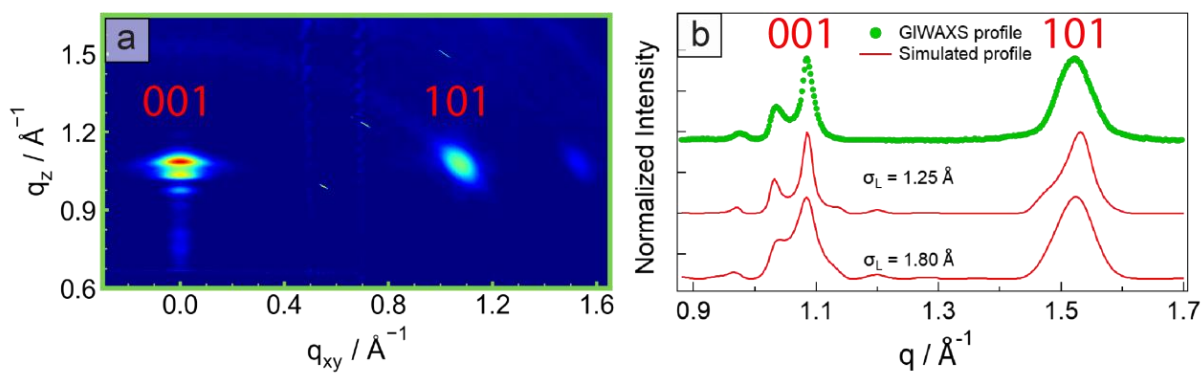

**Figure S9.** (a) 2D GIWAXS pattern of C$_{10}$ superlattice and (b) comparison between the 001 and 101 profiles extracted from (a) and the calculated diffraction patterns of C$_{10}$ superlattice with $\sigma_L \approx 1.28$ and 1.8 Å.

**Table S3. Multilayer diffraction fit results of Bragg patterns.** Structure parameters extracted from the multilayer diffraction fits. Fits are shown in Figure S10.

| Sample | d [Å] | L [Å] | $\sigma_L$ [Å] | N [planes] | $\sigma_N$ [planes] |
|---|---|---|---|---|---|
| **(100) Out-of-plane** | | | | | |
| **5 nm-C$_{18}$ NCs** | 5.879 | 36.483 | 2.040 | 8.831 | 1.366 |
| **8 nm-C$_{18}$ NCs** | 5.828 | 36.359 | 1.510 | 12.743 | 1.998 |
| **C$_{12}$** | 5.835 | 36.912 | 0.866 | 12.762 | 0.746 |
| **C$_{10}$** | 5.808 | 32.905 | 0.958 | 13.778 | 0.573 |
| **C$_8$** | 5.808 | 32.901 | 1.203 | 19.116 | 3.501 |
| **(101) Out-of-plane** | | | | | |
| **C$_{10}$** | 4.108 | 26.477 | 1.853 | ≈16 | n/a |
| **C$_8$** | 4.103 | 29.994 | 1.264 | ≈20 | n/a |

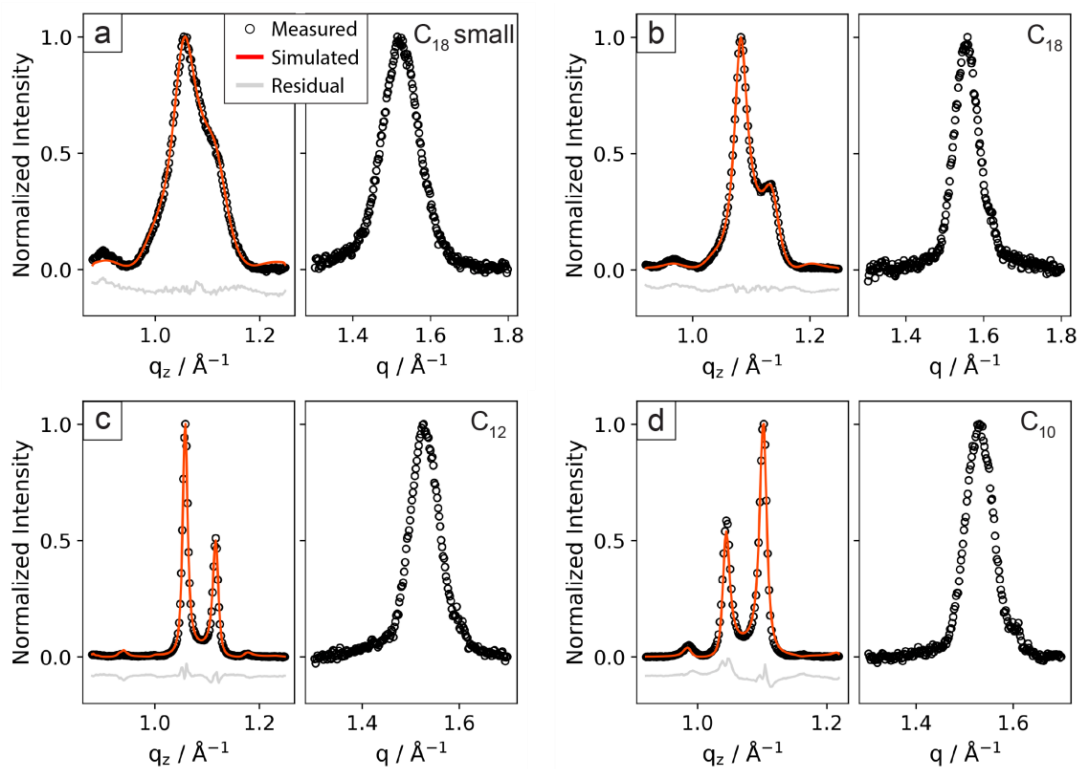

**Figure S10. Multilayer diffraction fits of Bragg patterns.** a-d) On the right are (100) out-of-plane peak profiles and on the left are (101) peak profiles extracted from the 2D GIWAXS patterns.

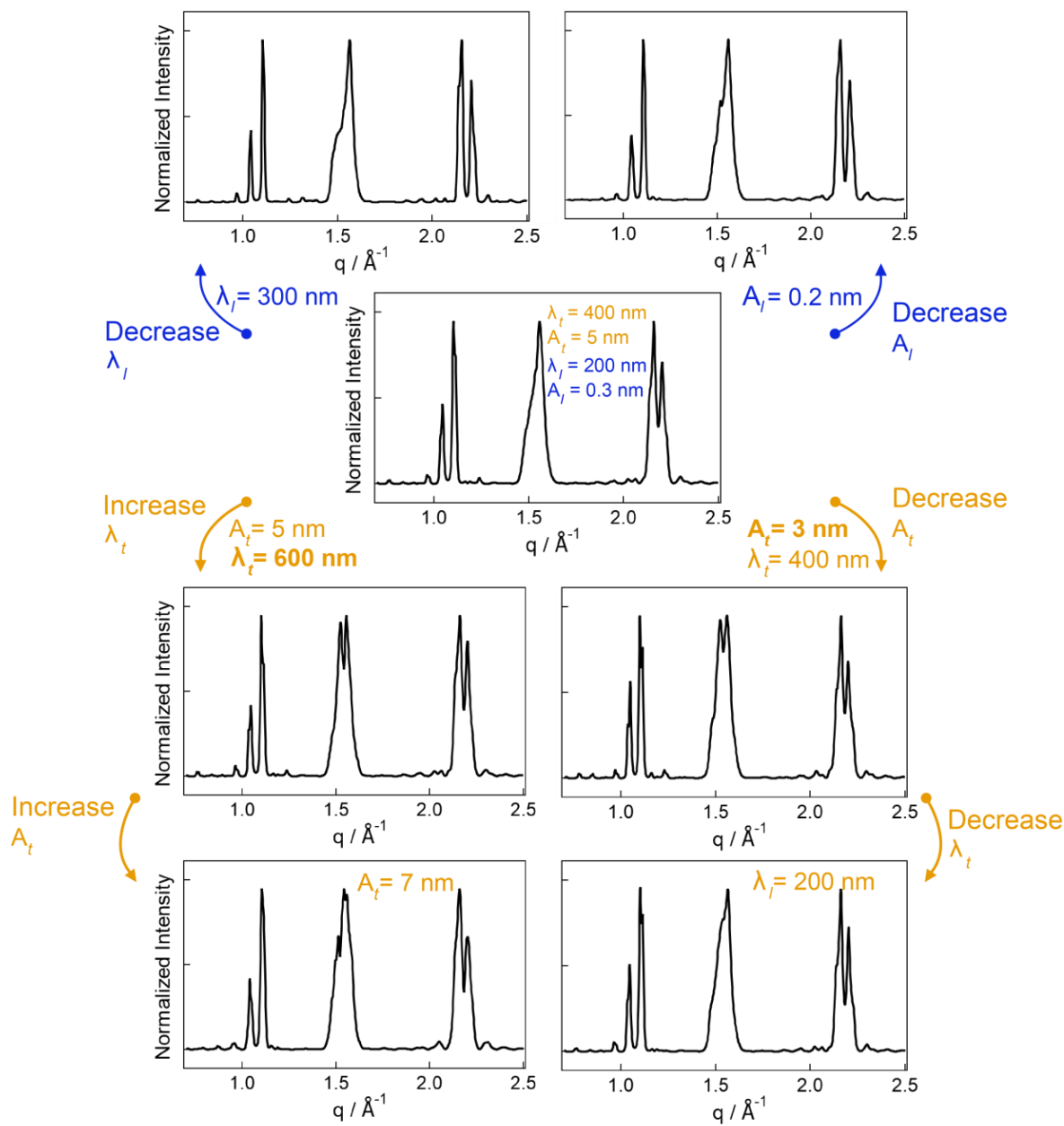

**Figure S11.** Simulated wide-angle X-ray diffraction patterns of CsPbBr$_3$ nanocrystal superlattices showing the effect of changing $\lambda_l$ (longitudinal wavelength), $A_l$ (longitudinal amplitude), $\lambda_t$ (transversal wavelength) and $A_t$ (transversal amplitude). The longitudinal parameter has little effect on the (101) peak, while it affects the broadening of the (001) and (002) peaks. As discussed in the main text, the broadening of the (101) peak is primarily influenced by the transversal parameters.